\pdfoutput=1
\documentclass[reprint,amsmath,amssymb,aps,prapplied,floatfix]{revtex4-2}
\makeatletter
\usepackage{listings}
\usepackage[hidelinks,colorlinks=true,linkcolor=blue,citecolor=blue, urlcolor = blue]{hyperref}
\usepackage{bookmark}
\newcommand*{\rom}[1]{\expandafter\@slowromancap\romannumeral #1@}
\usepackage{parskip}
\usepackage{graphicx}
\usepackage{dcolumn}
\usepackage{comment}
\usepackage{bm}
\usepackage[utf8]{inputenc}
\usepackage{lineno}
\begin{document}
\preprint{APS/123-QED}

\title{Updated Trends in Neutrino-induced hadron production}
\author {Rashi Sharma} 
\email{	rsharm18@syr.edu}
\affiliation{Department of Physics, Syracuse University, Syracuse, NY 13244, USA}

\author {R. Aggarwal}
\email{ritu.aggarwal1@gmail.com}
\affiliation{USAR, Guru Gobind Singh Indraprastha University, East Delhi Campus, 110092, India}

\author {M. Kaur}
\email{manjit@pu.ac.in}
\affiliation{Department of Physics, Panjab University, Chandigarh 160014, India\\
Department of Physics, Amity University, Punjab, Mohali {140306}, India}

\date{\today}

\begin{abstract}

With four different type of neutrino-induced interactions, we considered to investigate and reanalyse the KNO scaling in modified multiplicity distributions from a different perspective.~In an attempt of first of its kind, we propose alternate fitting function to parameterise the distribution than the most widely adopted Slattery's function and compare it with yet another form.~We propose the shifted Gompertz and Weibull functions as the fitting functions and compare their potency for the most conventional form of Slattery's function.~In addition the analysis of the data by evaluating the central moments and factorial moments, we show the dependence of moments on the target size. 

\end{abstract}

\keywords{Probability distributions, Shifted Gompertz and the Weibull functions, KNO scaling, normalised Moments.}
\maketitle


\section{\label{sec:level1} INTRODUCTION}
Study of multiplicity distributions of charged hadrons produced in lepton-induced and hadron-induced interactions in different targets has remained in focus ever since the advent of high energy and cosmic ray physics.~It has been extensively studied in fixed-target and collider experiments as well as in cosmic ray experiments.~The results of such studies are utilised in modeling of interaction dynamics.~In contrast to the vast information available from experiments using leptons and hadrons as probes, accessibility of such information from neutrino-induced experiments has remained very limited.~Earliest studies on the charged hadronic multiplicities in charged-current~(CC) and neutral-current~(NC) interactions measured in experiments performed with 15-foot bubble chamber during 1970s to the latest results from the OPERA experiment using CERN CNGS neutrino beams, and the CHORUS experiment have provided results in different center-of-mass (cms) energies and in different phase space regions \cite{ Ope,Ope2,Ope3,kuzmin,Chor}.
The mean charged-hadron multiplicities in the muon-neutrino and muon-antineutrino charged-current reactions on hydrogen and deuterium have been measured in the Fermilab experiments E31 \cite{PhysRevD.17.1,new}, E45 \cite{PhysRevLett.36.124,PhysRevD.19.1} and E-545 \cite{KITAGAKI1980325,zie} with the 15-foot Bubble Chamber and in the CERN experiments WA21 \cite{jones1990w,jones1992} and WA25 \cite{WA25:1983qub,WA25:1984qub} with the Big European Bubble Chamber (BEBC). The data obtained with the FNAL and BNL hydrogen bubble chambers before 1976 are gathered in Ref. \cite{Albini:1975iu}.
\\

The motivation for the present study stems from the fact that there is one common investigation which has been performed on the data from all these experiments.~It relates to the study of Koba-Nielsen-Olesen~(KNO) scaling \cite{Koba:1972ng,Gazdzicki:1990bp}, a study which provides understanding of improving models of particle production which are used in Monte Carlo (MC) event generators. 
~In almost all the results on KNO distributions, different experiments using neutrino beams have used the Slattery's function in different forms, to fit the KNO distributions \cite{Ope, Chor, zie}
However, the multiplicity data in neutrino interactions are very rare. Only recently two emulsion based neutrino experiments OPERA and CHORUS  \cite{Ope,Chor} published multiplicity distributions and also tested KNO scaling with a reliable statistics. The multiplicity distribution for each data follows a negative binomial distribution exhibiting approximate KNO scaling.~The KNO scaled distribution has been fitted with the Slattery's function.

The aim of the present work is to show that the KNO distribution can be defined in terms of different functions with improved precision than the Slattery's function.~A comparison of two different distributions, namely the shifted Gompertz and the Weibull distributions with the Slattery's distribution is presented.~In addition we also evaluate  central and factorial moments of the multiplicity distributions, both from the data and the best fitted proposed distribution.~With four type of neutrino interactions considered, we set out to study the effect of KNO scaling in their multiplicity distributions.~Two new fitting functions are proposed and their potency for all the above cases is studied.

~One of the consequences of the KNO scaling is that the dispersion over mean multiplicity is rendered independent of kinematic quantities.~The probability distribution of $n$-particle events is also well represented by the moments of the distribution and its generating function. The analysis of multiplicity moments is a powerful tool which helps to unfold the characteristics of the multiplicity distribution.~Calculated as derivatives of the generating function, the particle correlations can be studied through the normalized central moments.~Dependence of moments on energy can be used to validate the KNO scaling \cite{Ansorge, Alner} or to check for violation.~Several analyses of multiplicity moments have been done at various energies, using different probability distribution functions~\cite{Suzuki,PhysRevD.54.4333,Capella1996,Prasz,Pandey}.~However, these analyses mostly are done for $e^{+}e^{-}$, $pp$  and $\overline{p}p$ collisions.~Such studies in neutrino-induced interactions are missing.~The presented work is the first analysis using new distributions for the case of $\nu$-X and $\overline{\nu}$-X interactions, where X is a target. 
\section{Methodology}
The multiplicity distribution is expressed in terms of the probability of producing $n$ number of particles in the final state of a collision.~The shape of such a distribution varies with system size and collision energy and can be incorporated into the study of its higher moments.
The multiplicity distribution of charged hadrons produced in the neutrino interactions reflects the characteristics of hadronic final states in hard scattering.~These type of data assist to improve models of particle production which are used in Monte Carlo~(MC) event generators.~The shape of the multiplicity distribution is often studied in terms of functional dependence of the probability on the number of particles $n$, produced in a collision.~The following sections describe various forms most commonly used and the new proposed functions.
\subsection{KNO Formalism}
Koba, Nielsen, and Olesen showed that when the multiplicity distributions were scaled by average multiplicity $\langle n_{ch}\rangle$, they became asymptotically independent of the energy of interaction. 
The KNO hypothesis shows that at very high center-of-mass (CMS) energy $\sqrt  s$, the probability $P_{n}$ of producing $n$ charged particles in a collision process having the mean number of charged particles $\langle n \rangle$, should follow the following scaling relation;
\begin{equation}
\small
P_{n}(s)=\frac{1}{\langle{n}\rangle}\psi(z,s)= \lim_{s \to \infty}  \frac{1}{\langle{n}\rangle}\psi(z), \hspace*{3mm}  {\rm where}  \hspace*{2mm}   z=\frac{n}{\langle{n}\rangle}, \label{eqKNO}
\end{equation}
Thus, the data points $P_{n}(s)$ measured at different energies $\sqrt{s}$ should fall on a single scaling curve defined by the function $\psi$.~This curve can then be parameterised by a fit-function.

\subsection{Parameterisation of KNO distributions}
Parameterisation of the KNO scaled distributions was first introduced by Slattery \cite{slatt} in the form;
\begin{equation}
\psi(z) = (A_{1}z^{3} + B_{1}z^{4})e^{-C_{1}z} \label{eq:slat}
\end{equation}
Various experiments  \cite{Chor, Ope, zie} used this form to fit the KNO distributions.
The data from the experiments involving neutrino interactions show that approximate KNO scaling as a function of an appropriate multiplicity variable $z{'}$ is valid for the charged hadrons multiplicity. However, the interaction energies are typically low, with $ W^{2}$ of the order of 35 GeV$^{2}$.~For $\nu_{\mu}$ charged-current~(CC) interactions;
\begin{equation}
W^{2} = 2m_{N}E_{had} + m_{N}^{2} - Q_{\nu}^{2},\label{eq:Wsq}\\    
\end{equation}

where $Q_{\nu}^{2}$ is the squared four-momentum transfer, and $m_{N}$ is the nucleon mass.~$W^{2}$ is the square of the invariant mass of the hadronic system.

The first observation of the KNO violation came from $pp$ interactions at the Intersecting Storage Ring (ISR).~The violation was soon discovered at other energies and in other interactions involving $e^{+}e^{-}$, $p\overline{p}$  etc.
KNO scaling violation led to the application of negative binomial distribution~(NBD), introduced by P. Carruthers et al \cite{Carruthers,Giovannini}.

With a low priority of using KNO scaled distributions, different experiments used NBD to unwind the mechanism of particle production.~The success of NBD was phenomenal in providing a description in a most consistent way.~Over a period of time, some more statistical distributions were introduced and used for interpreting the data.~These include Gamma distribution \cite{Urmossy}, Lognormal distribution \cite{Genesis}, Tsallis distribution \cite{Tsallis, Tsallis2} and the more recent Weibull distribution (Wei)~\cite{W1951, Sadhana}.~Nevertheless, at very high energies, typically in the TeV range, NBD was also seen to deviate.

In the present work we introduce a yet novel way to parameterise the KNO distribution and show that its agreement is far improved in comparison to the Slattery's function. We choose shifted Gompertz distribution~(SGD) to fit the KNO scaled distributions and compare it with Slattery's function and the Weibull distribution.~A description of this new distribution follows in the next section.

\subsection{ The shifted Gompertz distribution~(SGD) }

In one of our earlier works, we put in place the use of shifted Gompertz distribution \cite{Chawla}, first introduced by \cite{Bemmaor1994}, to investigate the multiplicities in leptonic and hadronic collisions for different collision energies.~The distribution interpreted the experimental data from high-energy particle collisions involving leptons and hadrons as probes, very well.

The SGD distribution uses two non-negative parameters; one of them is known as the scale parameter and the other a shape parameter.~Taking these parameters as $b>0$ and $t>0$, the probability density function of a variable $n$ is then defined as;
\begin{equation}
P_{n} = b e^{-bn}e^{-\big(t e^{-bn}\big)}\big[1+t(1-e^{-bn}\big)\big]\hspace{0.5cm} n > 0 , \label{SGD}
\end{equation}
Maximum of two independent random variables with Gompertz distribution (parameters $b > 0$ and $t > 0$) and an exponential distribution (parameter $b > 0$), characterise the distribution.~We used SGD in describing multiplicity data in $e^+e^-$, $e^{+}p$, $pp$, $\bar{p} p$ data at different energies and showed that SGD provides a good description \cite{Chawla, singla, ritu, PhysRevD.102.054005}.

\subsection{ The Weibull distribution}

A highly versatile probability density function~(pdf), the Weibull distribution \cite{wei} can fit a wide range continuous data.~It has been used to study the data from different regimes such as medicine, quality control, engineering etc. and can also be used to model the skewed data quite well.~The probability density function of this distribution is expressed in three different forms; 3-parameter Weibull, 2-parameter Weibull and 1-parameter Weibull. \\
The 3-parameter Weibull probability density function is given by;
\begin{equation}
    P_{n} = \frac{\beta}{\eta}\left( \frac{n-\gamma}{\eta}\right )^{\beta -1}e^{-(\frac{n-\gamma}{\eta})^{\beta}}
\label{eq7}
\end{equation}

where $P_{n} \geq 0$, $n > \gamma$, where $\gamma$ is the location parameter and  $-\infty < \gamma < +\infty$. The shape parameter $\beta >0$  and the scale parameter $\eta >0$.\\ By setting $\gamma$=0, one gets the 2-parameter Weibull pdf;
\begin{equation}
    P_n = \frac{\beta}{\eta}\left( \frac{n}{\eta}\right )^{\beta -1}e^{-(\frac{n}{\eta})^{\beta}}
\label{eq:Wei2}
\end{equation}
The 1-parameter Weibull assumes the only unknown as the scale parameter $\eta$, the shape parameter $\beta$ is known a-priori and hence a constant, with equation (\ref{eq:Wei2}) is used and compared with results from SGD.

\subsection{Moments of multiplicity distribution}
The moments are calculated as derivatives of the generating function and the particle correlations can be studied through the normalised moments ($C_{q}$) and normalised factorial moments ($F_{q}$) which are defined as \cite{Mangeol}; 
\small
\begin{gather}
C_{q} = \frac{\langle n^q\rangle}{\langle n\rangle ^q} = \frac{\sum_{n}{n^{q}P_{n}}}{(\sum_{n} nP_{n})^q} \label{eq:Cq}\\
F_{q} = \frac{\sum_{n=q}^{n_{\it max}}n(n-1).......(n-q+1)P_{n}}{(\sum_{n=1}^{n_{\it max}}nP_{n})^q}\label{eq:Fq}\
\end{gather}
\normalsize

\subsection{Trends in mean multiplicity dependence on \texorpdfstring{$W^{2}$}{TEXT} }
In order to compare the dependency of average charged hadron multiplicity on the invariant mass of hadronic system, the data from different experiments are studied.~The invariant hadronic mass is expressed as in equation (\ref{eq:Wsq}).~The mean charged hadron multiplicity is found to vary linearly as a function of logarithm of the square of the invariant mass of the hadronic system $W$, in various ranges of $W^2$,
 \begin{equation}\label{eq:1}
 \langle n_{ch} \rangle = a + b(lnW^2) 
 \end{equation}
 This linear variation was found to be true for all energies.

 \subsection{Dispersion trends}
 Dispersion $D$ (= $\sqrt {\langle n_{ch}^{2}\rangle - \langle n_{ch}\rangle^{2}})$ of multiplicity distribution of $n$ particles is interesting from theoretical point of view.~It is understood that for independent particle emission, the dispersion versus average multiplicity should follow a Poisson distribution.~However, it is observed that in hadronic interactions the variation of dispersion follows an empirical relation with multiplicity as;
\begin{equation} \label{eq:2}
D = A + B \langle n_{ch} \rangle
\end{equation} 
For data from different experiments, we study this dependence.~The interpolation of the fit from equation (\ref{eq:2}) gives an unexpected intercept at the $\langle n_{ch} \rangle$ axis.~The value of the intercept is used to modify the KNO distribution for improving fitting with different functions.

\section{Data analysed}
Data from the four major experiments have been analysed with details as given below:
\subsection{Data from the OPERA experiment}
The OPERA experiment was designed to observe and study the neutrino oscillations in the $\nu_{\mu}$$\rightarrow$$\nu_{\tau}$ oscillations in appearance mode in the CNGS (CERN Neutrinos to Gran Sasso) neutrino beam \cite{Autiero:2009zz, Acquafredda:2009}.~The experiment established neutrino oscillations with the discovery of $\nu_{\tau}$ appearance with a significance of 5.1$\sigma$ \cite{Ope}.
~The OPERA detector was a hybrid setup consisting of electronic detectors and a massive lead-emulsion target.~The nuclear emulsions were used as very precise tracking devices and electronic detectors to locate the neutrino interaction events in the emulsions.~It was exposed to the CNGS $\nu_{\mu}$ beam with mean energy of 17 GeV. A data sample corresponding to 1.8×10$^{20}$ protons on target~(p.o.t.) collected during the period 2008 to 2012 as published in \cite{Acquafredda:2009}), the electronic detectors recorded 19505 neutrino interactions in the target fiducial volume.~A sub-sample of 818 events occurring in the lead with a negatively charged muon was selected in order to measure the track and vertex parameters in the target including a detailed check of the nuclear break-up and evaporation processes.~Imposing an additional requirement of selecting events with $W^{2}> $1 GeV$^{2}$ to eliminate quasi-elastic events, a total of 795 events were selected.~Description of the OPERA detector and selection procedures can be found in \cite{Acquafredda:2009}.

~In the present analysis, we have used the charged hadron multiplicities obtained from 795 $\nu_{\mu}$-$Pb$ events in different $W^{2}$ ranges and corrected for efficiencies from the paper by N. Agafonova et al \cite{Ope, Ope2, Ope3}

\subsection{Data from the CHORUS experiment}
The CHORUS experiment was designed to search for $\nu_{\mu}\rightarrow \nu_{\tau}$ oscillations.~The CHORUS hybrid detector was exposed to the wide band neutrino beam of the CERN SPS during the years 1994–1997, with an integrated flux of 5.06 $\times 10^{19}$ protons on nuclear emulsion target.~The West Area Neutrino Facility (WANF) of the CERN SPS provided an intense beam of neutrinos with an average energy of 27 GeV.~More than $10^{6}$ neutrino interactions were accumulated in the emulsion target.~ A requirement on the square of the invariant mass of the hadronic system, $W^{2} >1$ GeV$^{2}$  along with other selection criteria to remove the background, was imposed.~A sample of 496 $\nu_{\mu}$-A and 369 ${{\overline {\nu}}_{\mu}}$-A events, A represents the target, was finally selected for analysis.~Details of the data and the efficiency corrections for every $W^{2}$ range can be obtained from the reference~\cite{Chor}.
~In the present work, we study the charged hadron multiplicities produced in these interactions, measured by the CHORUS collaboration.~Investigation into the KNO scaling \cite{Koba:1972ng} behaviour of the charged hadron multiplicity in different kinematical regions has been done. 

\subsection{{\texorpdfstring{$\nu n$}{}} and {\texorpdfstring{$\nu p$}{}} charged-current interactions from Fermilab Bubbble Chamber}
Charged-hadron multiplicity distributions in $\nu n$ and $\nu p$ charged$-$current~(CC) interactions were measured in an exposure of the Fermilab deuterium-filled 15-foot bubble chamber to a wide-band neutrino beam produced by 350-GeV protons.~Charged-hadron multiplicities initiated in charged-current neutrino interactions on deuteron targets, from which $\nu n$ and $\nu p$ collisions from an identical neutrino flux were separated.
The data sample corresponds to a flux of $4.57\times10^{18}$ protons on target. The average neutrino energy was 50 GeV.
 \begin{gather}
 \nu_{\mu} + n \rightarrow \mu^{-}+X^{+},\hspace{0.3cm} X^{+}\rightarrow hadrons\\
 \nu_{\mu} + p \rightarrow \mu^{-}+X^{++}, \hspace{0.2cm} X^{++}\rightarrow hadrons
 \end{gather}
~Charged hadron multiplicity distributions published in \cite{zie} measured for a) 9237 neutrino-neutron CC interactions and b) 6033 neutrino-proton CC interactions, distributed over different $W^{2}$ ranges between 1-225 GeV$^2$ are used for the present analysis.~The paper by D. Zieminska et al \cite{zie}, contains details of the data and selection procedure used.\\  

\subsection{{\texorpdfstring{$\nu p$}{}} charged current interactions from Fermilab Bubble Chamber}
The multiplicity distributions of the hadrons produced in antineutrino-proton interactions in a sample consisting of 2025 charged-current events with
antineutrino energy greater than 5 GeV are analysed. The data~\cite{new} comes from exposures of the 15-foot hydrogen
bubble chamber to the broad-band antineutrino beam at Fermilab.~The distribution in hadronic mass $W$ has an average value of 3.7 GeV but extends up to 10 GeV.~The data samples were obtained from three separate exposures of the Fermilab 15-foot hydrogen bubble chamber.~The events were obtained with a 400~GeV proton beam incident on an aluminium target.~Two horns were used to focus the produced negative particles which in turn decayed to generate the  ${{\overline {\nu}}_{\mu}}$ beam. 
The ${{\overline{\nu}p}}$ charged-current~(CC) events were extracted from the sample that included contributions from both CC and NC reactions.

$\overline{\nu}p\rightarrow\mu^{+}H^{0},\\
\nu p\rightarrow \mu^{-}H^{++},\\
\overline{\nu}p\rightarrow {\overline\nu}H^{+},\\
\nu p\rightarrow \nu H^{+}$

The bulk of the charged-current data are in the
$W$ range, $2<W <6$ GeV with a median $W$ value of 3.7 GeV.
The details of the data and the efficiency corrections for every $W$ range can be obtained from the reference~\cite{new}.
\section{RESULTS}
\begin{figure}[!ht]
\includegraphics[scale=0.48]{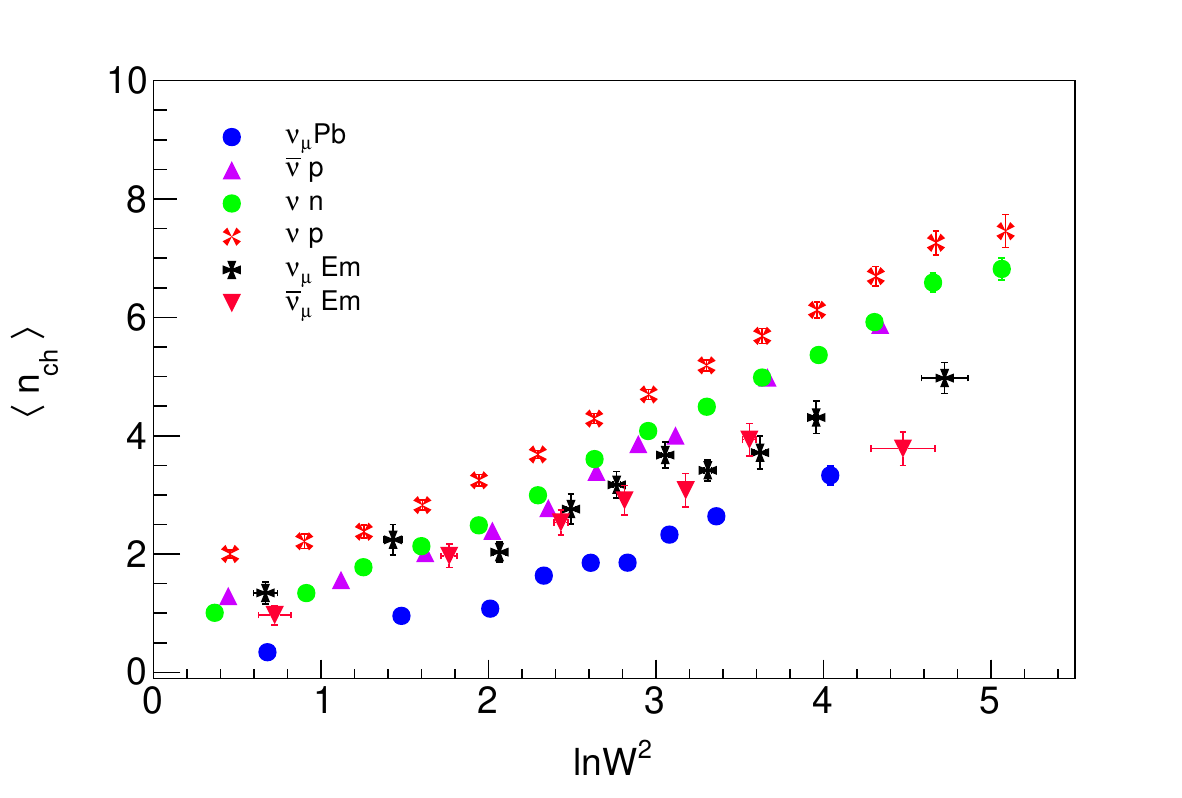}
\caption{Average charged hadron multiplicity $\langle n_{ch}\rangle$ as a function of $ln$$ W^2$.~The data on interactions i) $\nu_{\mu}$-Pb collected by the OPERA experiment using CERN-CNGS \cite{Ope} ii) $\nu$-n and $\nu$-p  obtained from FNAL-Bubble Chamber \cite{zie}iii)  $\overline{\nu}$-p using FNAL-Bubble Chamber and \cite{new} iv) $\nu_{\mu}$-Em and  $\overline{\nu}_{\mu}$-Em by the CHORUS experiment \cite{Chor}.}  
\label{fig:nch_vs_lnW2_All}
\end{figure}

Figure~\ref{fig:nch_vs_lnW2_All} shows the dependence of $\langle n_{ch}\rangle$ on $lnW^2 $ for each specified data being analysed.~Most of the earlier studies made a linear fit, $\langle n_{ch}\rangle = a + b ln W^{2}$ to each data set.~Accordingly a linear fit has been made to validate the data used, and the values of parameters $a$ and $b$ are found to be very close to the earlier results, as shown in Table \ref{table:lnW}. 

\begin{table}[ht]
\caption{ Linear fit parameters for $\langle n_{ch} \rangle$ versus $lnW^2$ dependence } 
\centering 
\begin{tabular}{c c c c c} 
\hline\hline 
Interaction & a & b & $\chi^2/ndf$ &  Ref. \\ [0.4ex] 
\hline 
$\nu_{\mu}$-Pb &-0.27 $\pm$ 0.06 & 0.81 $\pm$ 0.03 & 22.55/7 & \cite{Ope}\\ 
$\nu_{\mu}$-Em & 0.69 $\pm$ 0.18 & 0.87 $\pm$ 0.06 & 11.8/8 & \cite{Chor}\\
$\bar{\nu}_{\mu}$-Em & 0.48 $\pm$ 0.18 & 0.84 $\pm$ 0.07 & 5.03/5 & \cite{Chor} \\
$\bar{\nu}$ p  & -0.40 $\pm$ 0.12 & 1.43 $\pm$ 0.05 & 12.6/6 & \cite{new}  \\
$\nu$ n & -0.19 $\pm$ 0.06 & 1.42 $\pm$ 0.02 & 14.4/9 & \cite{zie}\\
$\nu$ p & 0.49 $\pm$ 0.13 & 1.43 $\pm$ 0.05 &  2.26/9 & \cite{zie} \\[1ex] 
\hline\hline 
\end{tabular}
\label{table:lnW} 
\end{table}

\begin{figure}[!ht]
\includegraphics[scale=0.42]{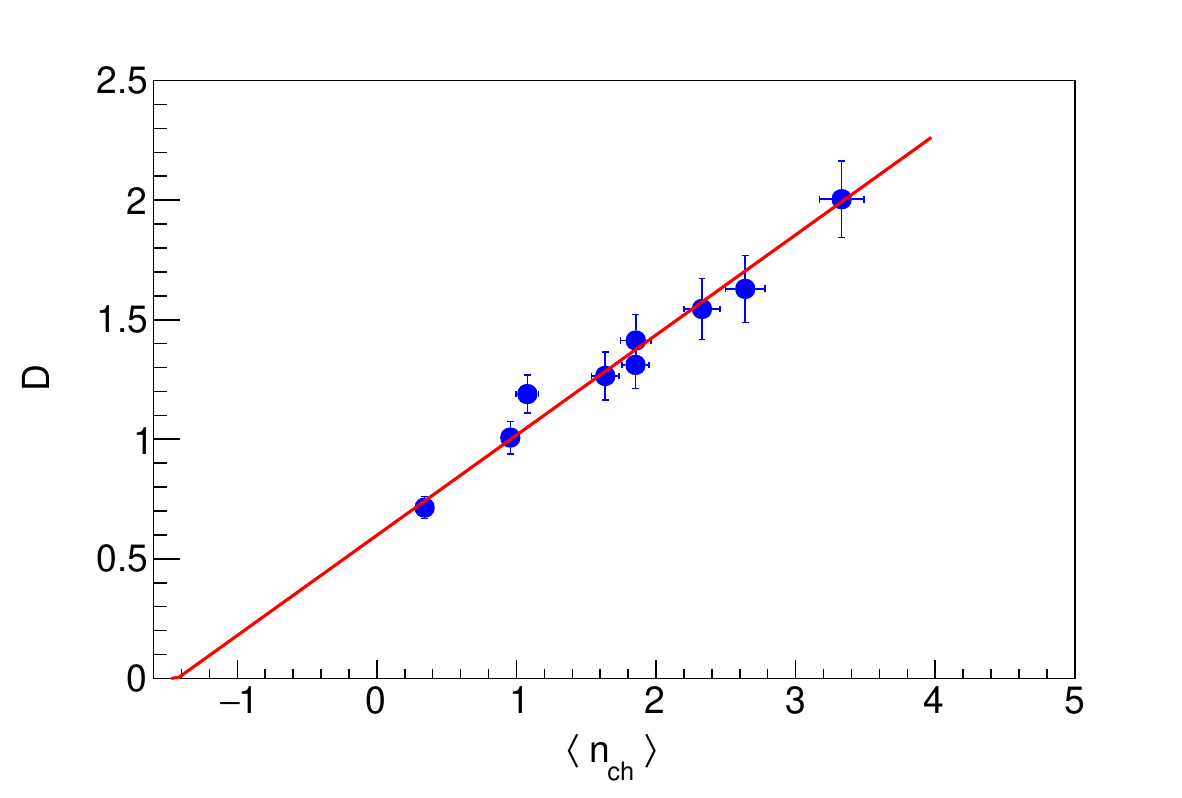}
\caption{Dispersion as a function of $\langle n_{ch}\rangle$  for $\nu_{\mu}$-Pb interactions obtained by the OPERA experiment \cite{Ope}.} 
\label{fig:D(n_ch)_vs_n_ch_C_Anu}
\end{figure}
Figure \ref{fig:D(n_ch)_vs_n_ch_C_Anu} shows the dependence of dispersion $D$ on the  average multiplicity $\langle n_{ch}\rangle$ for data on $\nu_{\mu}$-Pb interactions, from the OPERA experiment.~A straight line fit equation(\ref{eq:2}) to the data confirms a linear dependence.~The interpolation of the straight line fit on the $\langle n_{ch}\rangle$ axis is measured as a parameter, $\alpha=-$A/B.~Similar linear dependencies are studied for all the data sets and the value of $\alpha$ obtained for each of the data sets.~To avoid multiple similar figures, only one of these, is presented here.~The fit coefficients A and B and the $\alpha$ values are shown in the Table \ref{table:Disp} for all the data sets being analysed.
\begin{table}
\caption{ Dispersion D versus $\langle n_{ch} \rangle$ variation.~Values of slope A, intercept $B$ of the linear fit and the ratio $\alpha=-$A/B are shown.} 
\centering 
\begin{tabular}{c c c c c c} 
\hline\hline 
Interaction & A & B & $\chi^2/ndf$ & $\alpha$ &  Ref. \\ [0.5ex] 
\hline 
$\nu_{\mu}$-Pb & 0.59 $\pm$ 0.05 & 0.42 $\pm$ 0.02  & 3.67/7  & -1.410 & \cite{Ope}  \\
$\nu_{\mu}$-Em & 1.37 $\pm$ 0.23 & 0.14 $\pm$ 0.07  & 8.11/8  & -6.468 & \cite{Chor}\\
$\bar{\nu}_{\mu}$-Em & 1.14 $\pm$ 0.23 & 0.25 $\pm$ 0.09 & 1.51/5 & -4.532 & \cite{Chor} \\
$\bar{\nu}$ p  & 0.54 $\pm$ 0.07 & 0.31 $\pm$ 0.03  & 2.39/8  & -1.723 & \cite{new}  \\
$\nu$ n        & 0.29 $\pm$ 0.04  & 0.36 $\pm$ 0.01 & 4.89/11 & -0.928 & \cite{zie}\\
$\nu$ p        & 0.09 $\pm$ 0.09 & 0.34 $\pm$ 0.02  &  2.26/9 & -0.299 & \cite{zie} \\
[1ex] 
\hline\hline 
\end{tabular}
\label{table:Disp} 
\end{table}
\subsection{Effect of {\texorpdfstring{$\alpha$}{}}}
\begin{figure}[!ht]
\includegraphics[scale=0.42]{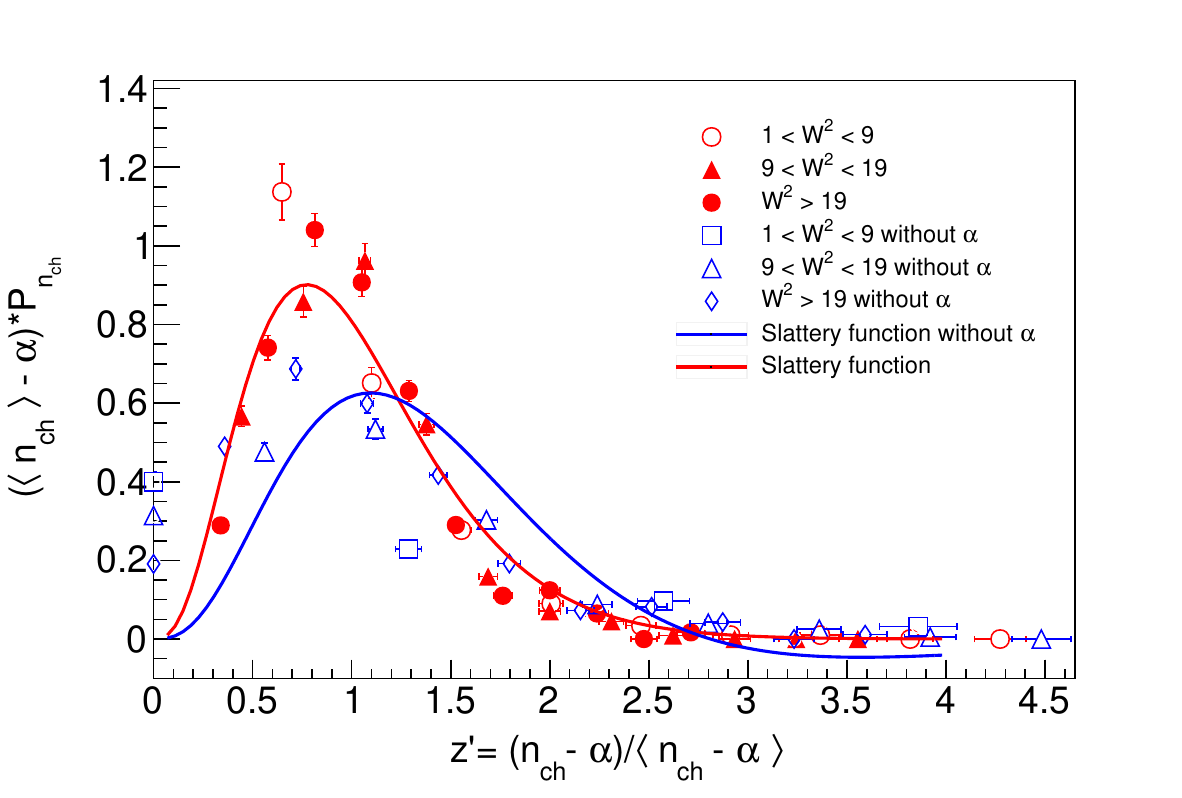}
\caption{KNO distributions in different $W$~(GeV) fitted with Slattery function i) without including $\alpha$~(blue solid line) and ii) with $\alpha$ included in the function~(red solid line) for the data from the OPERA experiment \cite{Ope}.} 
\label{fig:OPP.pdf}
\end{figure}

KNO scaling as discussed in section II was derived from Feynman scaling, observing that at high energy the KNO leads to an asymptotic scaling of the total multiplicity as $\langle n_{ch}\rangle$ $\propto$ ln$\sqrt{s}$.~ Thus, KNO scaling implies that the intercept A in equation(\ref{eq:2}) be compatible with 0, which is not the case at low to medium energies for all kinds of interactions.~Alpha is calculated from equation (\ref{eq:2}) as $\alpha = -$A/B, a variable which is reaction independent but energy dependent.~Using $\alpha$, Buras et al \cite{buras1973} provided an extension of the KNO scaling to low energies by introducing a new variable $z{'}$ defined as:
\begin{equation} z{'}= \frac{n_{ch}-\alpha}{\langle n_{ch}-\alpha\rangle} \label{eq:zp}
\end{equation}
A possible explanation of $\alpha$ has been proposed in terms of a leading particle effect in interactions using hadrons, neutrinos as well as resulting from the heavy nuclear targets in experiments using nuclear emulsions \cite{buras1973, Baranov1983, Chor}. Figure \ref{fig:OPP.pdf} shows two KNO distributions for $\psi(z=\frac{n_{ch}}{\langle n_{ch}\rangle})$ and $\psi(z{'}=\frac{n_{ch}-\alpha}{\langle n_{ch}-\alpha\rangle})$ for the data from the OPERA experiment.~The distributions are fitted with the Slattery's function, equation(\ref{eq:slat}), Shifted Gompertz function, equation(\ref{SGD}) and the Weibull function, equation(\ref{eq:Wei2}). From Table~\ref{table:compare}, it is observed that $\chi^{2}/ndf$ falls by a large factor when including $\alpha$ to calculate $z{'}$.~This shows a big enhancement in performance of the fit functions and justify the use of $\alpha$ to modify the variable $z$ to $z{'}$.~The same trend is observed for all the data under study.

\begin{table}
\caption{KNO distribution fitted with different functions, with and without $\alpha$ for the OPERA data \cite{Ope}.~Ratio of $\chi^{2}$ values w.r.t. fit with $\alpha$ are shown.} 

\centering 
\begin{tabular}{c c c c } 
\hline\hline 
Interaction & Fit-distribution & $(\chi^2_{\alpha}/ndf)$   &   $(\chi^2/ndf)$\\ 
\hline 
$\nu_{\mu}$-Pb & Slattery & 1 & 19.05 \\
$\nu_{\mu}$-Pb & SGD& 1 & 7.20  \\ 
$\nu_{\mu}$-Pb & Weibull & 1 & 5.96\\ 
\hline\hline 
\end{tabular}
\label{table:compare} 
\end{table}

\subsection{Different probability distribution functions}
The KNO distribution has been studied by almost all the high energy physics experiments.~For the case of neutrino interactions various forms of Slattery's function \cite{slatt} introduced in 1973, have been used to fit the distribution.~Analysis of the data from the OPERA and the CHORUS experiments, used the function in the forms;
\begin{gather}
\psi(z{'}) = (Az{'}^{3} + Bz{'}^{4})e^{-Cz{'}} \label{eq:slatP1}\\
\psi(z{'}) = (Az{'}+Bz{'}^{3}-Cz{'}^{5}+Dz{'}^{7})e^{-Ez{'}} \label{eq:slatP2}
\end{gather}
where A, B, C, D, E are the fit parameters.

In section II(B-D) we discussed the Slattery's function for KNO distribution, Shifted Gompertz distribution and Weibull distributions.~Details of the distributions are also provided.~In the present work, we apply these distributions to investigate the KNO distributions with respect to the earlier used Slattery's function.

\begin{figure}[!ht]
\includegraphics[scale=0.42]{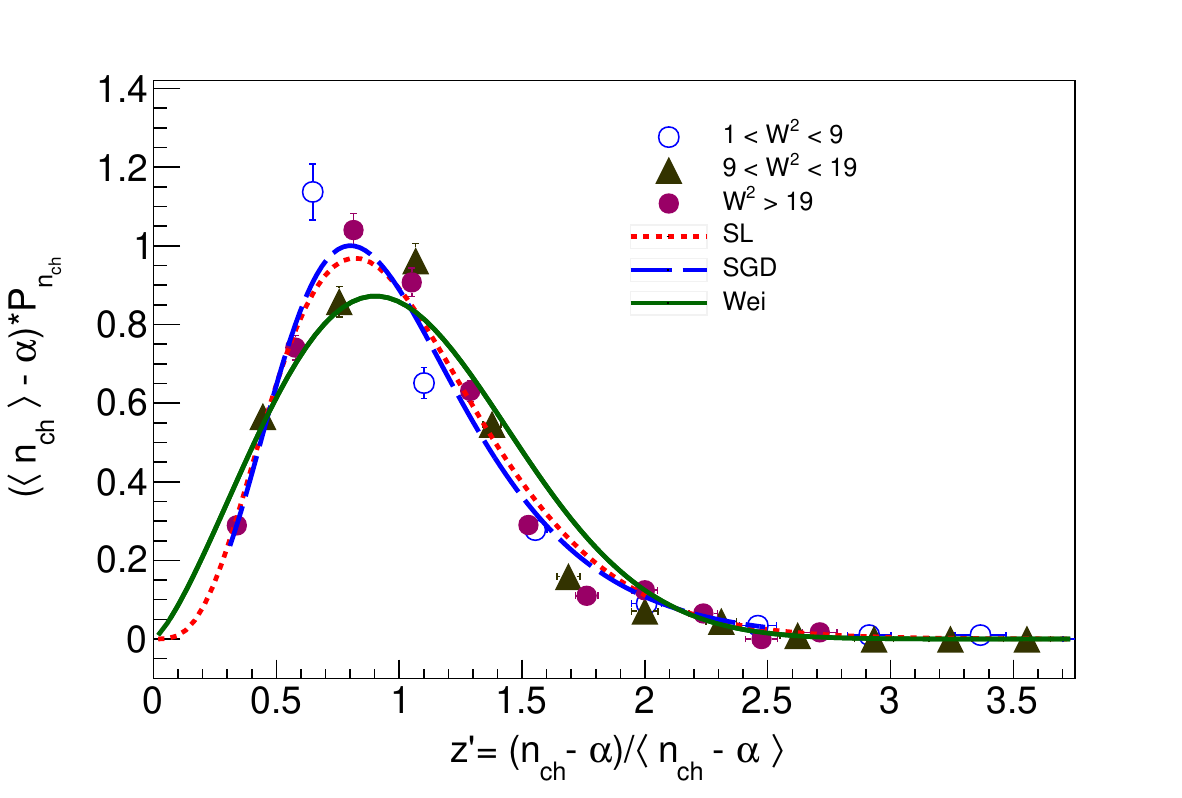}
\caption{KNO distributions of $\nu_{\mu}-$Pb data from the OPERA experiment \cite{Ope}, in different $W$~(GeV) ranges, fitted with three different functions.} 
\label{fig:All_FIT_OP}
\end{figure}

Figure \ref{fig:All_FIT_OP} shows the KNO distribution for the $\nu_{\mu}$-Pb interactions in three $W$~(GeV) ranges with $1<W^{2}<9$, $9<W^{2}<19$ and $W^{2}>19$ GeV$^{2}$ from the data obtained by the OPERA experiment \cite{Ope}.

\begin{figure}[!ht]
\includegraphics[scale=0.42]{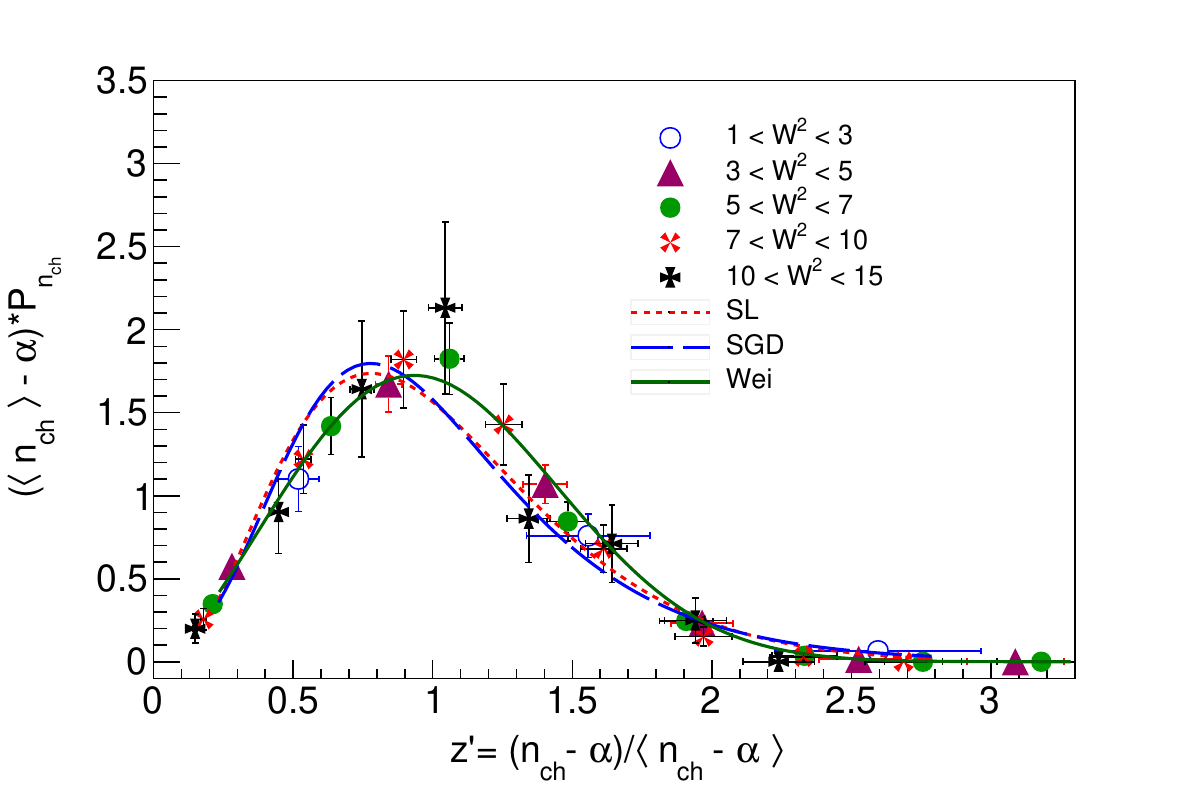}
\caption{KNO distributions of $\nu$-n data from reference~\cite{zie}, in different $W$(GeV) ranges, fitted with three different functions.} 
\label{fig:All_FIT_Zie_N_New}
\end{figure}

\begin{figure}[!ht]
\includegraphics[scale=0.42]{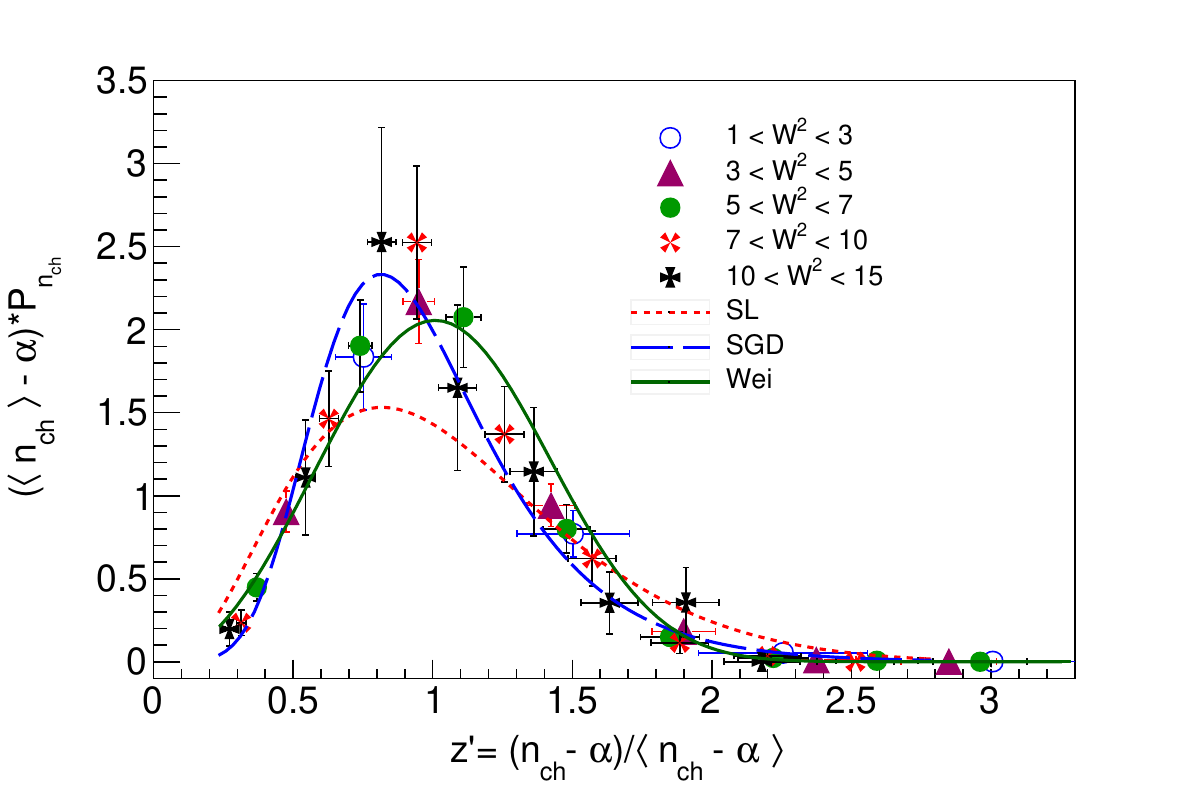}
\caption{KNO distributions of $\nu$-p data from reference~\cite{zie}, in different $W$(GeV) ranges, fitted with three different functions.} 
\label{fig:All_FIT_Zie_P_New}
\end{figure}

Figures \ref{fig:All_FIT_Zie_N_New} and \ref{fig:All_FIT_Zie_P_New} show the KNO distributions for the $\nu$-n and $\nu$-p interactions in five $W$~(GeV) ranges with $1<W^{2}<3$, $3<W^{2}<5$ and $5<W^{2}<7$, $7<W^{2}<10$ and $10<W^{2}<15$ GeV$^{2}$ for the data from reference \cite{zie}.

\begin{figure}[!ht]
\includegraphics[scale=0.42]{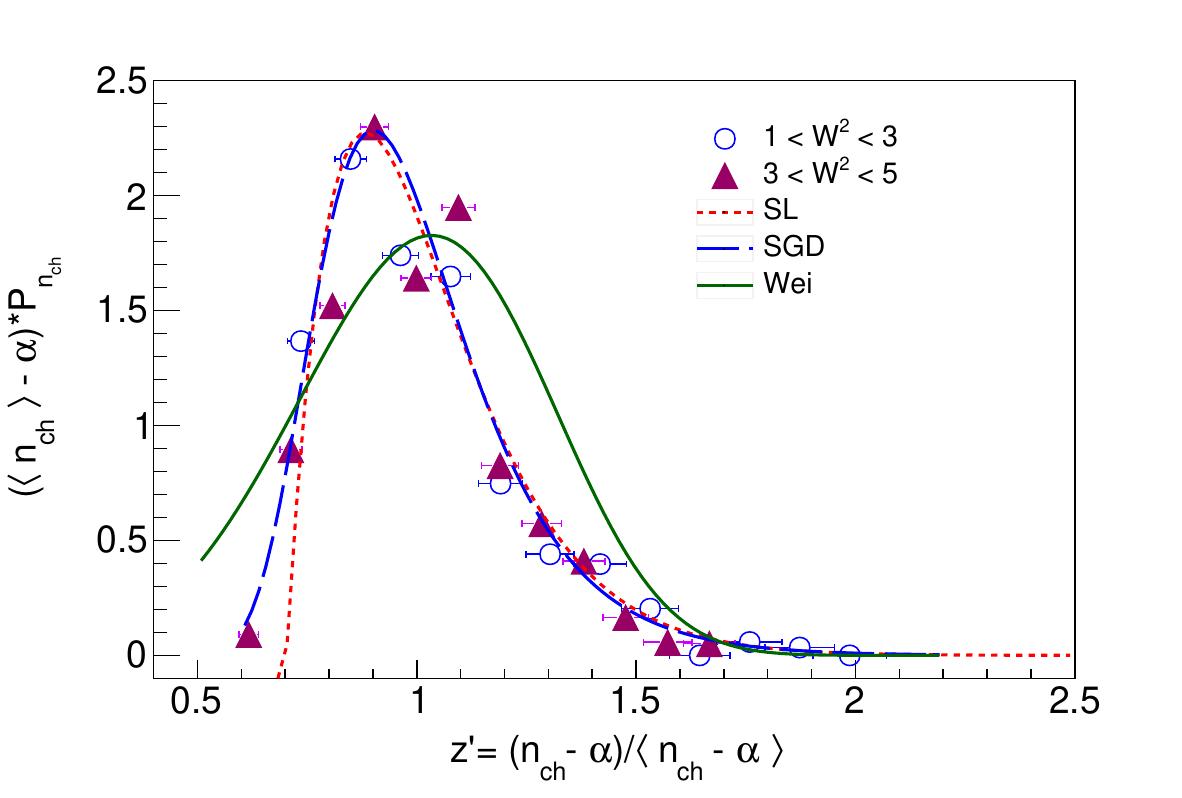}
\caption{KNO distributions of $\nu_{\mu}$-Emulsion data from the CHORUS experiment~\cite{Chor}, in different $W$(GeV) ranges, fitted with three different functions.} 
\label{fig:All_FIT_Chor}
\end{figure}

\begin{figure}[!ht]
\includegraphics[scale=0.42]{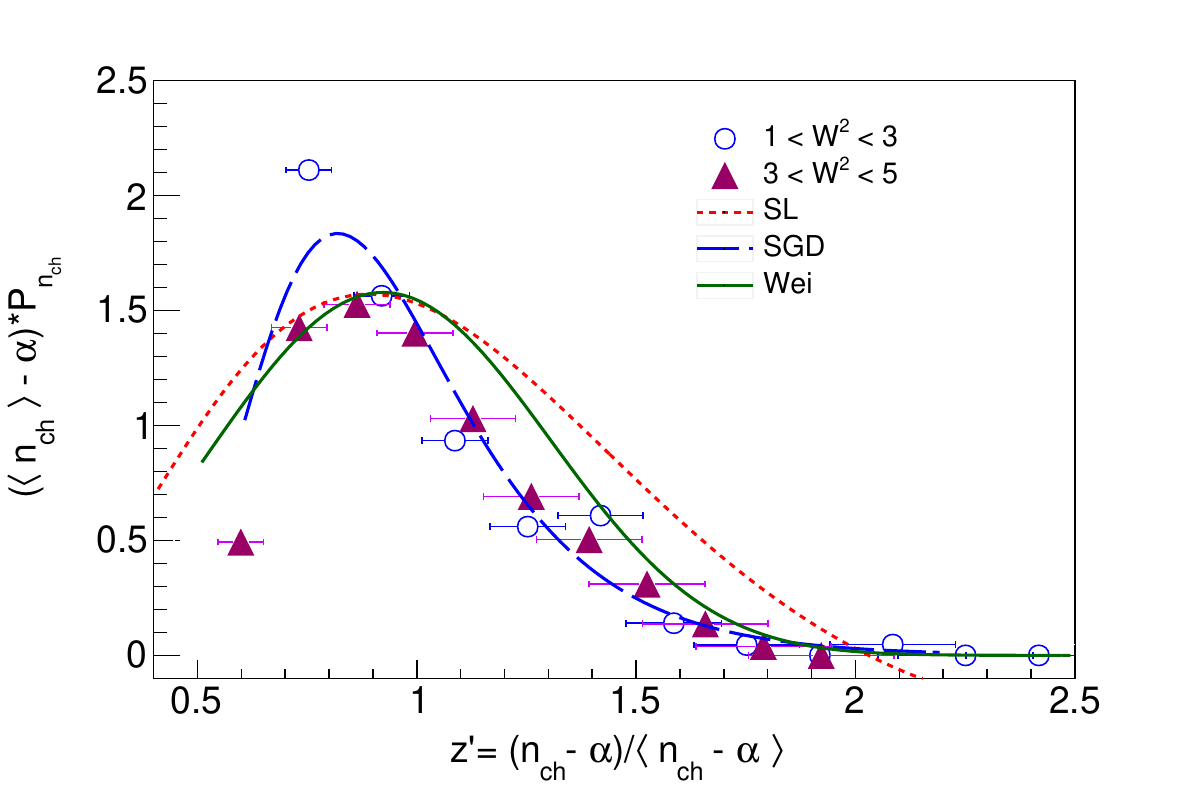}
\caption{KNO distributions of $\bar{\nu}_{\mu}$-Emulsion data from the CHORUS experiment~\cite{Chor}, in different $W$(GeV) ranges, fitted with three different functions.} 
\label{fig:All_FIT_Cho}
\end{figure}

Figures \ref{fig:All_FIT_Chor},\ref{fig:All_FIT_Cho} show the KNO distributions for the $\nu_{\mu}$-Emulsion and $\bar{\nu}_{\mu}$-Emulsion interactions in two $W$~(GeV) ranges with $1<W^{2}<3$ and $3<W^{2}<5$ GeV$^{2}$ for the data from the CHORUS experiment \cite{Chor}.

\begin{figure}[!ht]
\includegraphics[scale=0.42]{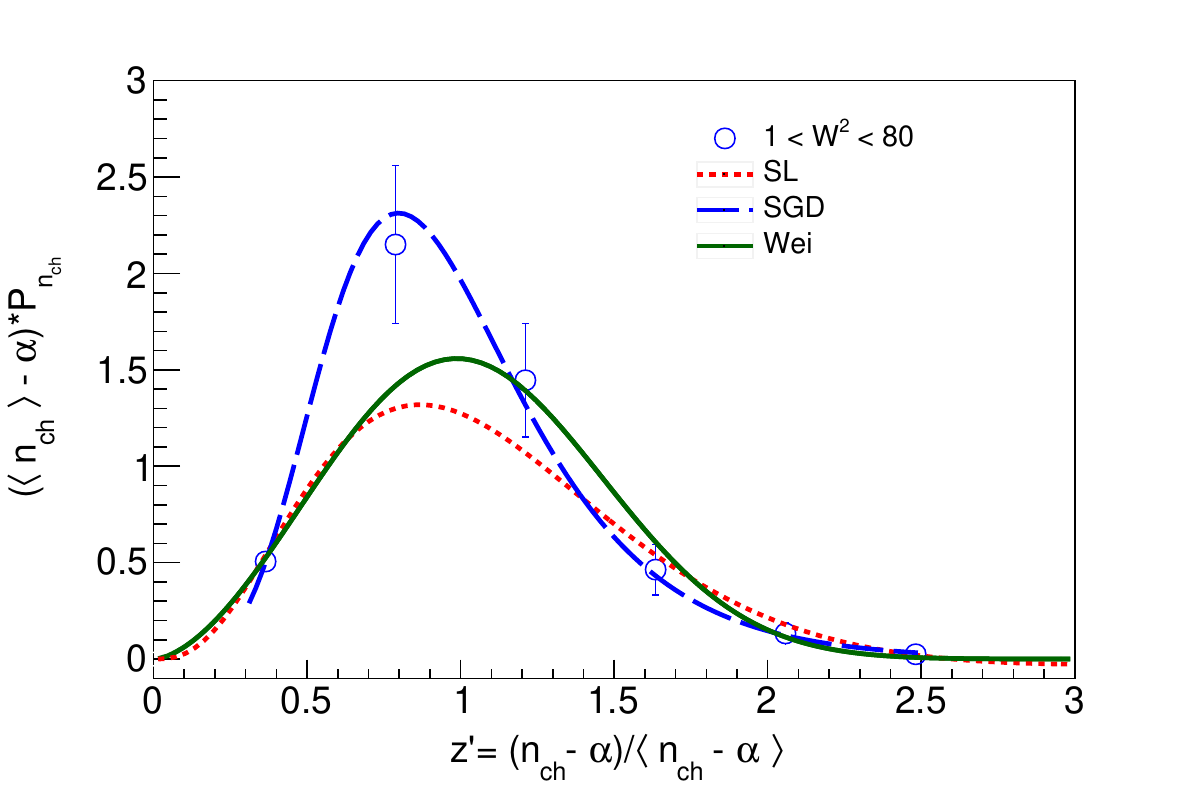}
\caption{KNO distributions of $\bar{\nu}$-proton data from reference~\cite{new} the with three different functions.} 
\label{fig:All_FIT_New}
\end{figure}

\begin{table}
\caption{Comparison of $\chi^{2}$ values for the three functions: SL(Slattery's ), SGD(Shifted Gompertz), Wei(Weibull's) fitted to the data.$^{*}$Square of the quoted $\langle W\rangle$.~$^{**}$The uncertainty in the determination of $W$ due to $\nu$-energy uncertainty is estimated to be 20 \% full width at half maximum(FWHM).} 
\centering 
\begin{tabular}{c c c c c c c} 
\hline\hline 
Interaction & $\langle W^{2}\rangle$ &    SL & SGD  & Wei & Ref. \\ [0.5ex] 
         &       GeV$^2$                 &      $\chi^{2}/ndf$& $\chi^{2}/ndf$& $\chi^{2}/ndf$& \\ 
\hline 
$\nu_{\mu}$-Pb & 16.9 $\pm$ 0.6 & 8.15 & 4.64 & 13.48 &  \cite{Ope} \\
$\nu_{\mu}$-Em & 17.7 $\pm$ 0.8 & 1.96 & 1.54 & 16.84 &  \cite{Chor}\\
$\bar{\nu}_{\mu}$-Em & 26.2 $\pm$1.3 & 1.74 & 0.24 & 0.79 &  \cite{Chor} \\
$\bar{\nu}$ p  & 28.62 $^{*}$ & 2.49 & 0.26 &  0.23 &  \cite{new} \\
$\nu$ n        & 28.03$^{**}$ & 0.71 & 0.93 & 0.27 &  \cite{zie}\\
$\nu$ p        & 29.41$^{**}$ & 2.65 & 1.11 & 0.61 &  \cite{zie} \\
[1ex] 
\hline \hline
\end{tabular}
\label{table:Chisq} 
\end{table}
The KNO distributed data in figures {\ref{fig:All_FIT_OP},\ref{fig:All_FIT_Zie_N_New},\ref{fig:All_FIT_Zie_P_New},\ref{fig:All_FIT_Chor},\ref{fig:All_FIT_Cho}}  are fitted with three distributions: i) Slattery's function, equation~(\ref{eq:slatP1}) ii) Shifted Gompertz function, equation~(\ref {SGD}) and iii) the Weibull function, equation~(\ref {eq:Wei2}).~Table \ref{table:Chisq} shows the $\chi^{2}/ndf$ values for each of the fits and for each data set.~It is observed that the Slattery's function gives the maximum $\chi^{2}/ndf$ for every data, thereby showing it to be a bad fit.~While the performance of both SGD and Weibull functions is highly improved in comparison to the Slattery's.~In addition, in all of the cases, the SGD turns out to be the best fit.
\subsection{Central and Factorial Moments}
Moment analysis is a powerful tool used for unfolding the characteristics of multiplicity distribution.~The moments are calculated as derivatives of the generating function of the probability distribution.~Higher factorial moments from which all other kinds of moments, factorial moments can also be calculated and the particle correlations can be studied through them.~The second central moment represents the variance of a random variable.~It captures how spread out a distribution is. Higher variance means a wider distribution.~The third moment called skewness, quantifies the relative size of the two tails of the distribution.~Skewness is negative for longer left tails and positive for longer right tails.~The third central moment is important because skewness is both location-and-scale-invariant.~The fourth central moment represents kurtosis  which is a measure of the combined size of the tails relative to whole distribution.~In a logical manner higher moments, odd-powered central moments quantify relative tailedness and even-powered moments quantify total peakedness. 
~Several analyses of moments have been done at different cms, using different probability distribution functions and variety of particles used as probes \cite{Prasz2,Capella1996,Suzuki,Pandey}.~The higher moments also can identify the correlations amongst particles produced in collisions.~Another study \cite{Rad} on evolution of the multiplicity distribution in a fireball that cools down after chemical freeze-out focused on to obtain different apparent temperatures from different moments.

\begin{figure}[!ht]
\includegraphics[scale=0.42]{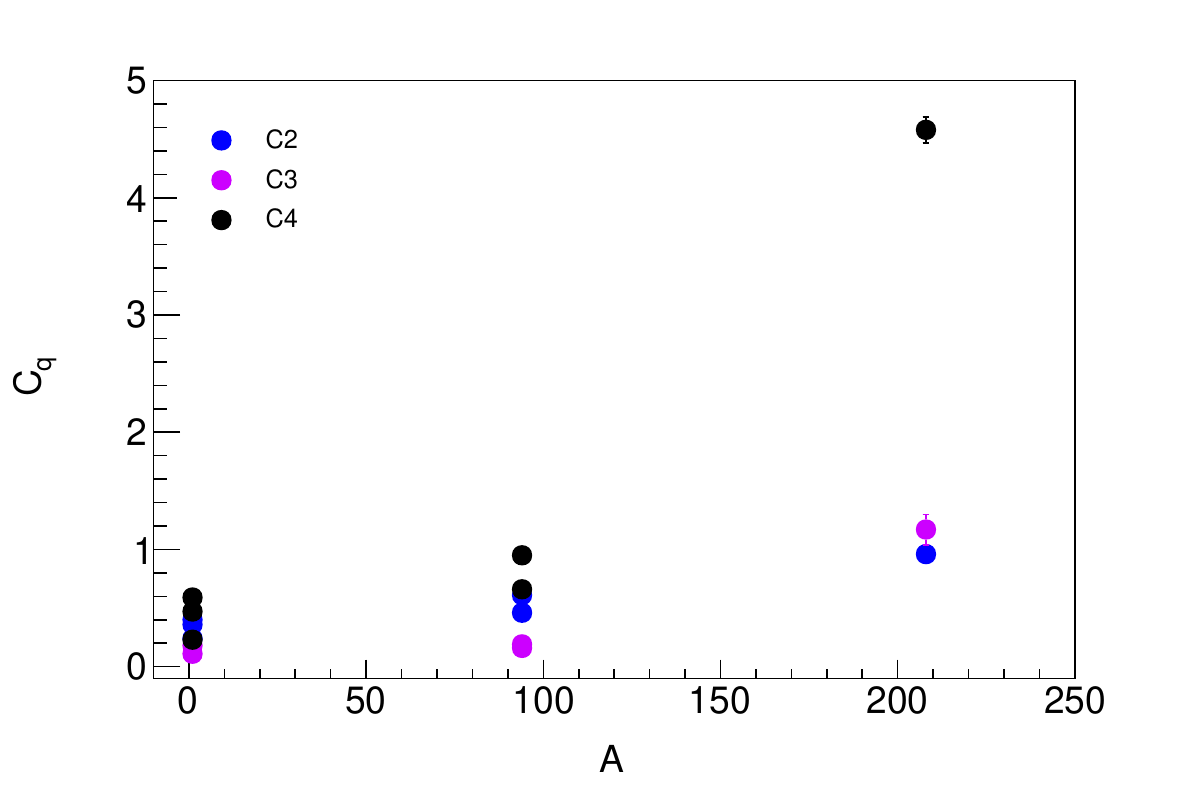}
\caption{Normalised central moments C\textsubscript{q} as a function of the target mass A in $\nu$-p~(A=1), $\nu$-n~(A=1), $\nu_{\mu}$-Em~(A=94) and $\nu_{\mu}$-Pb(A=207) interactions, obtained from the data \cite{Ope,Chor,zie}.} 
\label{fig:C5}
\end{figure}

\begin{figure}[!ht]
\includegraphics[scale=0.42]{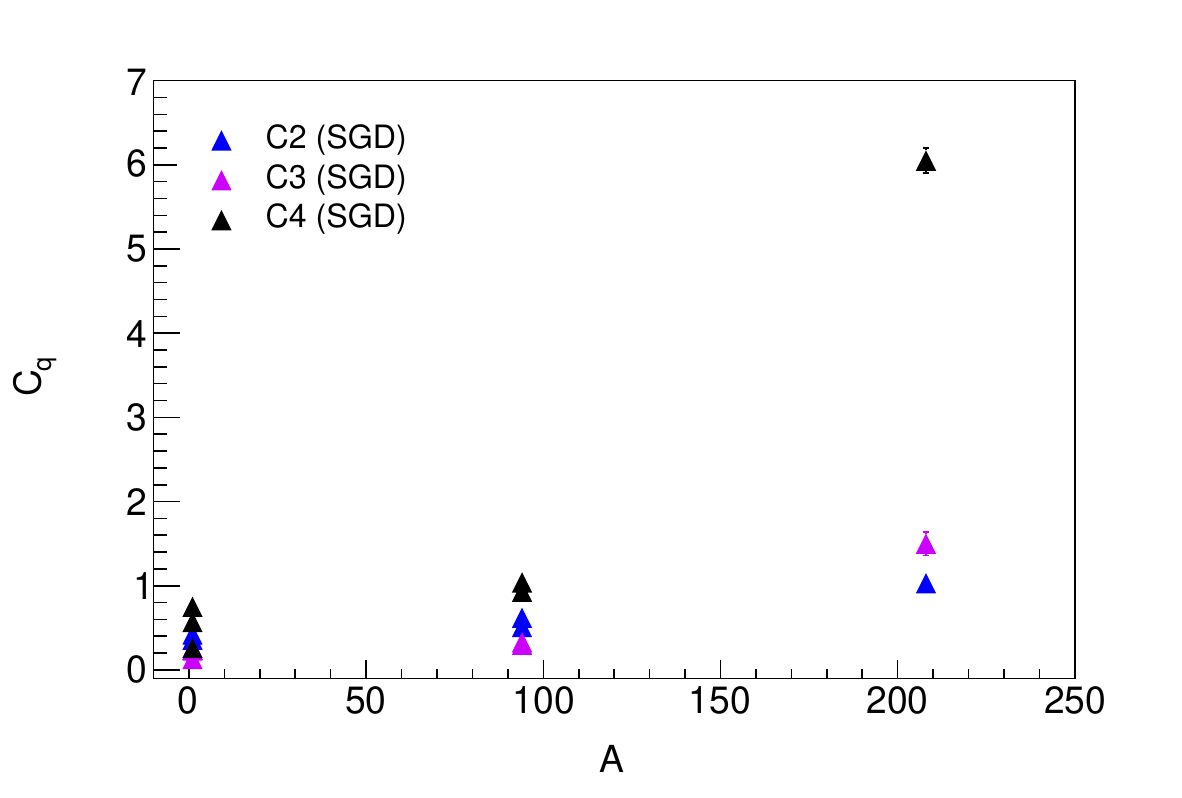}
\caption{Normalised central moments C\textsubscript{q} as a function of the target mass A in $\nu$-p~(A=1), $\nu$-n~(A=1), $\nu_{\mu}$-Em~(A=94) and $\nu_{\mu}$-Pb(A=207) interactions, obtained from the SGD fit to the data \cite{Ope,Chor,zie}.} 
\label{fig:C5S}
\end{figure}

\begin{figure}[!ht]
\includegraphics[scale=0.42]{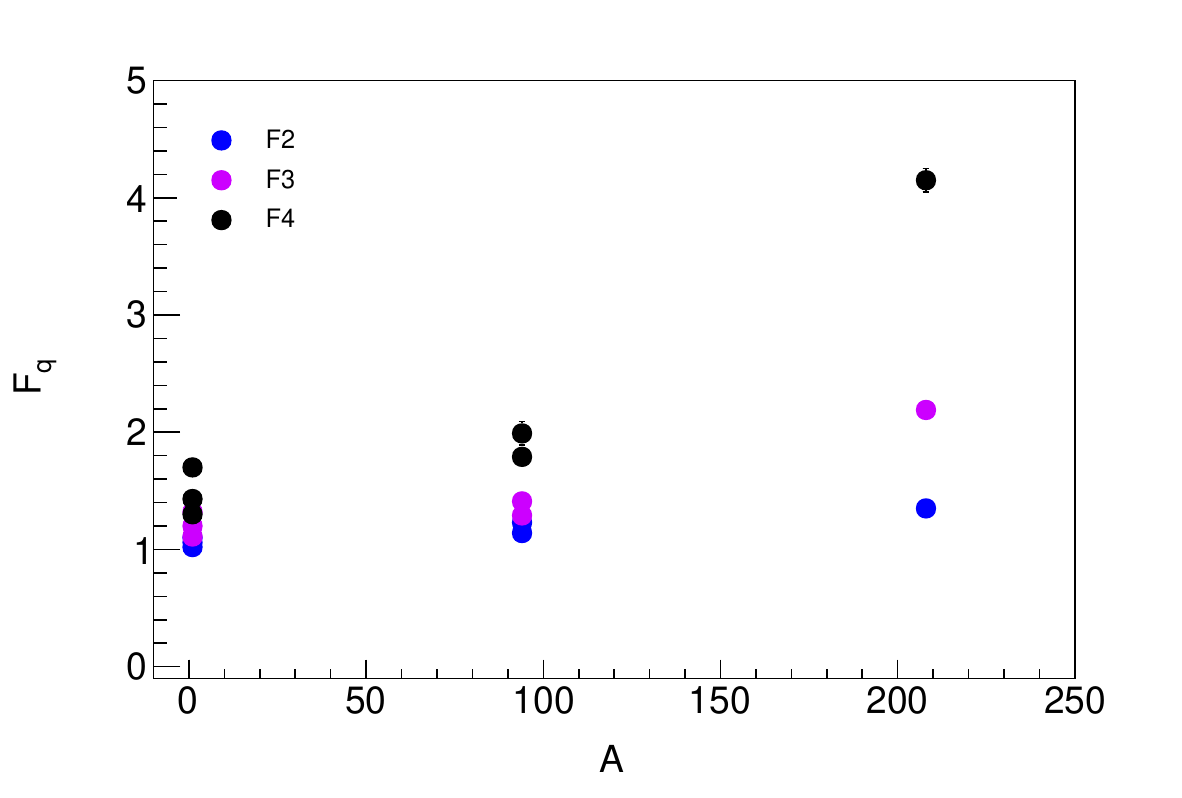}
\caption{Normalised factorial moments F\textsubscript{q} as a function of the target mass A in $\nu$-p~(A=1), $\nu$-n~(A=1), $\nu_{\mu}$-Em~(A=94) and $\nu_{\mu}$-Pb(A=207) interactions, obtained from the data \cite{Ope,Chor,zie}.}  
\label{fig:F5}
\end{figure}

\begin{figure}[!ht]
\includegraphics[scale=0.42]{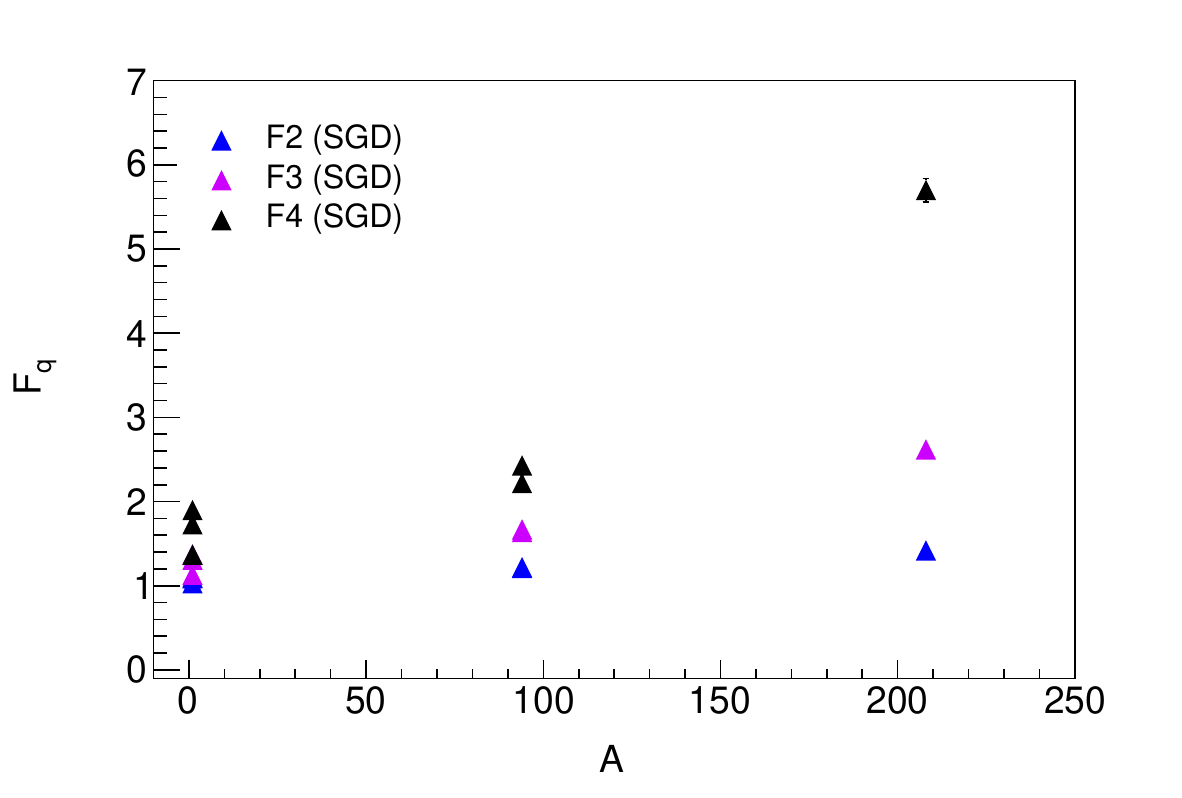}
\caption{Normalised factorial moments F\textsubscript{q} as a function of the target mass A in $\nu$-p~(A=1), $\nu$-n~(A=1), $\nu_{\mu}$-Em~(A=94) and $\nu_{\mu}$-Pb(A=207) interactions, obtained from the SGD fit to the data \cite{Ope,Chor,zie}.} 
\label{fig:F5S}
\end{figure}

 The values of central moments C\textsubscript{q} and factorial moments F\textsubscript{q} calculated for the experimental data and the SGD distributions which are the best fits of the data, are given in the Tables \ref{table:SGD fitted C_q} and \ref{table:SGD fitted F_q}.~Central moments are computed in terms of deviations from the mean, because then the higher-order central moments relate only to the spread and shape of the distribution.~It may be observed that the normalized central moments as well as normalized factorial moments obtained from the shifted Gompertz distribution are in good agreement with the experimental values.~This serves as a good test of the validity of the proposed SGD distribution.~Additionally, it is also observed from the $\overline{\nu}(\nu)$-Emulsion interactions and $\overline{\nu}$$(\nu)$-proton interactions that all moments have higher values for the case of $\overline{\nu}$ interactions than the corresponding $\nu$ interactions with the same target.~It is also observed that the values of both the central and factorial moments depend upon the target size A.~Figures \ref{fig:C5},\ref{fig:C5S} and  show the variation of C\textsubscript{2}, C\textsubscript{3}, C\textsubscript{4} moments derived from the experimental and SGD distributions. Similarly Figures \ref{fig:F5},\ref{fig:F5S} show the variation of F\textsubscript{2}, F\textsubscript{3}, F\textsubscript{4} moments derived from the experimental and SGD distributions.~From these figures and the Tables \ref{table:SGD fitted C_q} and \ref{table:SGD fitted F_q} it is found that the moments rise very fast for neutrino interactions with the target size; proton/neutron (A=1) to Emulsion (A=94) to Lead (A=207).

\begin{table*}
\small
\setlength{\tabcolsep}{0.35\tabcolsep}
\caption{Normalized central moments C\textsubscript{q} of experimental and shifted Gompertz distributions.} 
\centering 
\begin{tabular}{c| c c c c |c c c c|c} 
\hline\hline 
 &&& Experimental&&&&SGD&&\\
\hline
Reaction &  C\textsubscript{2} & C\textsubscript{3} & C\textsubscript{4} & C\textsubscript{5} & C\textsubscript{2} & C\textsubscript{3} & C\textsubscript{4} & C\textsubscript{5} & Ref \\ [0.3ex] 
\hline 
$\nu_{\mu}$-Pb &    0.96 $\pm$ 0.02 & 1.17 $\pm$ 0.13 & 4.58 $\pm$ 0.11 & 14.11 $\pm$ 0.35      & 1.03 $\pm$ 0.03 & 1.50 $\pm$ 0.14 & 6.05 $\pm$ 0.15 & 21.08 $\pm$ 0.52 &\cite{Ope}\\ 
$\bar{\nu}$-Em &  0.61 $\pm$ 0.03 & 0.16 $\pm$ 0.01 & 0.95 $\pm$ 0.05 & 0.92 $\pm$ 0.05     & 0.62 $\pm$ 0.03 & 0.30 $\pm$ 0.01 & 1.04 $\pm$ 0.05 & 1.31 $\pm$ 0.07 &\cite{Chor}\\
$\nu$-Em  & 0.46 $\pm$ 0.02  & 0.19 $\pm$ 0.01 & 0.66 $\pm$ 0.03 & 0.78 $\pm$ 0.03    & 0.51 $\pm$ 0.02 & 0.33 $\pm$ 0.01 & 0.93 $\pm$ 0.04 & 1.40 $\pm$ 0.05  &\cite{Chor}\\
$\bar{\nu}$-p & 0.40 $\pm$ 0.01 & 0.18 $\pm$ 0.01 & 0.59 $\pm$ 0.01 & 0.71 $\pm$ 0.01 & 0.42 $\pm$ 0.01 & 0.26 $\pm$ 0.01 & 0.75 $\pm$ 0.014 & 1.10 $\pm$ 0.02&\cite{new}  \\
$\nu$-n  & 0.36 $\pm$ 0.01 & 0.18 $\pm$ 0.02 & 0.47 $\pm$ 0.01 & 0.64 $\pm$ 0.01      & 0.36 $\pm$ 0.01 & 0.23 $\pm$ 0.01 & 0.57 $\pm$ 0.01 & 0.88 $\pm$ 0.01  &\cite{zie}\\
$\nu$-p & 0.24 $\pm$ 0.01 & 0.11 $\pm$ 0.01 & 0.23 $\pm$ 0.01 & 0.30 $\pm$ 0.01      & 0.24 $\pm$ 0.01 & 0.13 $\pm$ 0.01 & 0.26 $\pm$ 0.01 & 0.35 $\pm$ 0.01&\cite{zie} \\
\hline \hline 
\end{tabular}
\label{table:SGD fitted C_q} 
\end{table*}	

\begin{table*}
\setlength{\tabcolsep}{0.35\tabcolsep}
\small
\caption{Normalized factorial moments F\textsubscript{q} of experimental and shifted Gompertz distributions. } 
\centering 
\begin{tabular}{c| c c c c| c c c c |c} 
\hline\hline 
 &&& Experimental &&&& SGD&&Ref.\\
 \hline
Reaction & F\textsubscript{2} & F\textsubscript{3} & F\textsubscript{4} & F\textsubscript{5} & F\textsubscript{2} & F\textsubscript{3} & F\textsubscript{4} & F\textsubscript{5}&  \\ [0.3ex] 
\hline 
$\nu_{\mu}$-Pb &    1.35 $\pm$ 0.03 & 2.19 $\pm$ 0.05 & 4.15 $\pm$ 0.10 & 8.30 $\pm$ 0.21 &              1.42 $\pm$ 0.04 & 2.62 $\pm$ 0.07 & 5.70 $\pm$ 0.14 & 13.38 $\pm$ 0.33 &\cite{Ope}\\
$\bar{\nu}$-Em &    1.23 $\pm$ 0.06 & 1.29 $\pm$ 0.08 & 1.99 $\pm$ 0.10 & 2.28 $\pm$ 0.11 &              1.22 $\pm$ 0.06 & 1.64 $\pm$ 0.08 & 2.22 $\pm$ 0.11 & 2.77 $\pm$ 0.14 &\cite{Chor}   \\
$\nu$-Em &          1.14 $\pm$ 0.04 & 1.41 $\pm$ 0.06 & 1.79 $\pm$ 0.07 & 2.26$\pm$ 0.09 &      1.21 $\pm$ 0.05 & 1.67 $\pm$ 0.07 & 2.43 $\pm$ 0.09 & 3.51 $\pm$ 0.14 &\cite{Chor}  \\
$\bar{\nu}$-p &     1.06 $\pm$ 0.02 & 1.20 $\pm$ 0.02 & 1.43 $\pm$ 0.03 & 1.72 $\pm$ 0.03 &      1.09 $\pm$ 0.02 & 1.31 $\pm$ 0.02 & 1.73 $\pm$ 0.03 & 2.30 $\pm$ 0.04  &\cite{new}\\
$\nu$-n   &         1.10 $\pm$ 0.02 & 1.32 $\pm$ 0.01 & 1.70 $\pm$ 0.02 & 2.31 $\pm$ 0.02 &   1.10 $\pm$ 0.01 & 1.37 $\pm$ 0.03 & 1.90 $\pm$ 0.02 & 2.79 $\pm$ 0.03 &\cite{zie} \\
$\nu$-p &           1.02 $\pm$ 0.01 & 1.11 $\pm$ 0.01 & 1.30 $\pm$ 0.02 & 1.62 $\pm$ 0.02 &  1.03 $\pm$ 0.01 & 1.13 $\pm$ 0.02 & 1.37 $\pm$ 0.02 & 1.76 $\pm$ 0.02 &\cite{zie}\\
\hline\hline 
\end{tabular}
\label{table:SGD fitted F_q} 
\end{table*}
\newpage
\section{CONCLUSION}
A detailed analysis of the neutrino interactions has been done using data from four different experiments.~This is the first study in which the KNO distribution is studied by using different functions than the conventional Slattery's function.~The shifted Gompertz distribution and the Weibull distributions are studied.~Both these distributions show a much better agreement with the data in comparison to the Slattery's function.~However, the shifted Gompertz distribution shows the best agreement out of the three.

The average multiplicity $\langle n_{ch}\rangle$ varies nearly linearly as a function of $ln W^2$, although at very low $W^2$ it slightly departs.~The dependence of the charged hadrons multiplicity on dispersion $D$ also follows a linear relation equation~($\ref{eq:2}$).

It is interesting to observe that the values of both the central and factorial moments depend upon the target size A.~Higher the atomic weight of the target, faster is the growth of the moments.~Additionally, it is also observed from the $\bar{\nu}(\nu)$-Emulsion interactions and $\bar{\nu}(\nu)$-proton interactions that all moments have higher values for the case of $\nu$ interactions than the corresponding $\bar{\nu}$ interactions with the same target.~A conclusive regularity in A-dependence can be studied if more number of data points with larger variation of target sizes becomes available.~The information dissemination from such an analysis, particularly using the higher moments, is often useful to study the patterns and correlations.\\

\section{ACKNOWLEDGEMENT}
Authors R. Sharma and R. Aggarwal acknowledge the support of the Department of Technology, Savitribai Phule Pune University, India, where some of the work published here was undertaken under the DST INSPIRE Faculty grant.

\bibliography{Num.bib}


\end{document}